Growth and transport properties under high pressure of PrOBiS$_2$ single crystals

Masanori Nagao[a,c*], Akira Miura[b], Ryo Matsumoto[c], Yuki Maruyama[a], Satoshi Watauchi[a], Yoshihiko Takano[c], Kiyoharu Tadanaga[b], and Isao Tanaka[a]

[a]*University of Yamanashi, 7-32 Miyamae, Kofu, Yamanashi 400-8511, Japan*

[b]*Hokkaido University, Kita-13 Nishi-8, Kita-ku, Sapporo, Hokkaido 060-8628, Japan*

[c]*National Institute for Materials Science, 1-2-1 Sengen, Tsukuba, Ibaraki 305-0047, Japan*

[*]Corresponding Author

Masanori Nagao

Postal address: University of Yamanashi, Center for Crystal Science and Technology

Miyamae 7-32, Kofu 400-8511, Japan

Telephone number: (+81)55-220-8610

Fax number: (+81)55-254-3035

E-mail address: mnagao@yamanashi.ac.jp

**Abstract**

PrOBiS$_2$ single crystals were successfully grown using a KCl flux. The obtained crystals had a plate-like shape with a typical size of 0.5–1.0 mm and well-developed *ab*-plane. The Pr valence of the grown crystals, as determined by an X-ray absorption fine structure spectroscopy analysis, was almost trivalent. The PrOBiS$_2$ single crystals did not exhibit superconductivity down to 0.25 K. The transport properties under high pressures up to 50 GPa ensured semiconducting behavior for 2–300 K using a diamond anvil cell.

**Main text**

**1. Introduction**

The superconductivity of $R$OBiS$_2$ ($R$: rare earth elements) have been intensively studied [1-5]. Superconductivity is triggered by carrier doping, which can be induced by anion doping, by substituting F at the O site. However, CeOBiS$_2$ without F-doping exhibits superconductivity due to the valence fluctuation between Ce$^{3+}$ and Ce$^{4+}$ [6,7]. F-free $R$OBiS$_2$ compounds are important for understanding the superconductivity mechanism; it is important to know the properties of the mother compounds even without superconductivity. The crystal growth of F-free $R$OBiS$_2$ with $R$=La, Ce, and (Ce$_{0.5}$Pr$_{0.5}$) have been reported previously [6-10]. However, F-free PrOBiS$_2$ single crystals have not been grown. Additionally, the superconductivity of F-free BiS$_2$-based materials is enhanced under high pressure [8].

In this paper, we successfully grew F-free PrOBiS$_2$ single crystals using a KCl flux. The obtained single crystals of PrOBiS$_2$ were characterized by X-ray absorption fine structure spectroscopy for the Pr valence, and the electrical transport properties down to 0.25 K, under high pressures up to 50 GPa.

**2. Experimental**

The PrOBiS$_2$ single crystals were grown by a high-temperature flux method [11–13]. The raw materials; Pr$_2$S$_3$ (99.9 wt%), Bi$_2$O$_3$ (99.9 wt%), and Bi$_2$S$_3$ (99.9 wt%), were weighed for a nominal composition of PrOBiS$_2$. The mixture of the raw materials (0.8 g) and KCl (99.5 wt%) flux (5.0 g) were ground using a mortar, and then sealed into an evacuated quartz tube. The quartz tube was heated at 1050 °C for 10 h, followed by cooling to 750 °C at a rate of 1 °C/h. The sample was then cooled to room temperature in the furnace. The resulting quartz tube was opened in air, and the obtained materials were washed and filtered by distilled water to remove the KCl flux.

The compositional ratio of the single crystals was evaluated by energy dispersive X-ray spectrometry (EDS) (Bruker, Quantax 70) associated with the observation of the microstructure using a scanning electron microscope (SEM) (Hitachi High-Technologies, TM3030). The obtained compositional values were normalized using S = 2.00, with the Pr, O, and Bi values measured to a precision of two decimal places. The identification and orientation of the grown crystals were performed by X-ray diffraction (XRD) using Rigaku MultiFlex with CuK$\alpha$ radiation. The valence state of the Praseodymium component in the grown crystals was estimated by an X-ray absorption fine structure (XAFS) spectroscopy analysis with an Aichi XAS beamline and synchrotron X-ray radiation (BL11S2: Experimental No.201801025). For the XAFS spectroscopy sample, the obtained single crystals were grinded and mixed with boron nitride (BN) powder, and pressed into a pellet with a 4 mm diameter and total weight of approximately 31 mg.

The resistivity–temperature ($\rho$–$T$) characteristics were measured by the standard four-probe method with a constant current density ($J$) mode and a physical property measurement system (Quantum Design; PPMS DynaCool). The electrical terminals were fabricated by Ag paste. The $\rho$–$T$ characteristics in the temperature range 0.25–15 K were measured with an adiabatic demagnetization refrigerator (ADR) option for PPMS. The magnetic field for the operation of the ADR, 3 T at 1.9 K, was applied; and subsequently removed. The temperature of the sample consequently decreased to approximately 0.25 K. The measurement of the $\rho$–$T$ characteristics was started at the lowest temperature (~0.25 K), which was spontaneously increased to approximately 15 K. The resistance under high pressure in the temperature range 2–300 K was measured by the four-probe method using a diamond anvil cell (DAC) with boron-doped diamond electrodes [14–16] on the bottom anvil of the nanopolycrystalline diamond [17]. Each electrode was at a distance of 5 μm with a 200 nm thickness. The plate-shaped crystal settled on the bottom, and therefore the resistance parallel to the flat-plane was measured. The cubic boron nitride powders with a ruby manometer were used as the pressure-transmitting medium. The applied pressure values were estimated by the fluorescence from the ruby powders [18]. The Raman spectrum from the culet on the top diamond anvil [19] was measured by an inVia Raman Microscope (RENISHAW). The resistance–temperature ($R$–$T$) characteristics with a constant current ($I$) mode under high pressure were measured by a physical property measurement system (Quantum Design: PPMS) with the DAC equipment.

## 3. Results and discussion

Figure 1 shows a typical SEM image of the PrOBiS$_2$ single crystal. The obtained single crystals had plate-like shapes with sizes and thickness in the ranges 0.5–1.0 mm and 10–30 μm, respectively. The estimated atomic ratio of the obtained single crystals was Pr:O:Bi:S=1.03:0.96:0.99:2.00, which agreed with the nearly stoichiometric ratio. On the other hand, K and Cl from the flux were not detected in the single crystals with a minimum sensitivity limit of approximately 1 wt%.

Figure 2 shows the XRD pattern of a well-developed plane in the single crystals. The presence of only the 00$l$ diffraction peaks, similar to F-doped PrOBiS$_2$ compound structures [13], indicated a well-developed *ab*-plane. The *c*-axis lattice constant was approximately 13.78(3) Å.

Figure 3 shows Pr $L_3$-edge absorption spectra of the $PrOBiS_2$ single crystals using an XAFS spectroscopy analysis at room temperature. The Pr $L_3$-edge of the $PrOBiS_2$ single crystals exhibited a peak at approximately 5968 eV, which was assigned to $Pr^{3+}$. This is consistent with the other XAFS spectroscopy results for the trivalent electronic configuration ($Pr^{3+}$) [20,21]. On the other hand, the peak position of the tetravalent electronic configuration ($Pr^{4+}$) is approximately 5979 eV, which was assigned to $Pr_6O_{11}$ [20,22]. We analyzed the Pr valence state in the $PrOBiS_2$ single crystals using a linear combination fitting from the $Pr_2S_3$ and $Pr_6O_{11}$ XAFS spectroscopy spectra. The ratios of the compounds $Pr_2S_3$ ($Pr^{3+}$) and $Pr_6O_{11}$ (one-third $Pr^{3+}$ and two-thirds $Pr^{4+}$) were 97 and 3%, respectively. Thus, $Pr^{4+}/(Pr^{3+}+Pr^{4+})$ was estimated at 2%. Therefore, the trivalent electronic structure was a major contributor, with only a small amount of the tetravalent electronic configuration.

Figure 4 shows the $\rho$–$T$ characteristics parallel to the *ab*-plane for the $PrOBiS_2$ single crystal. The electrical resistivity decreased with decreasing temperature, indicating metallic behavior in the range 50–300 K. In contrast, semiconducting behavior occurred below 50 K, and the reason for this is unclear. Further investigation is required to clarify the origin of this phenomenon. The $\rho$–$T$ characteristics for the $PrOBiS_2$ single crystal in the temperature range 0.25–15 K had a non-zero resistivity, indicating that there were no superconducting transitions down to 0.25 K. This result is consistent with poly-crystalline samples [6].

An X-ray photoelectron spectroscopy analysis of the F-doped $PrOBiS_2$ ($PrO_{1-x}F_xBiS_2$: $x$ = 0.13 and 0.23) single crystals showed that the Pr valence state was almost three [23]. This valence was close to that obtained for $PrOBiS_2$, although the XAFS spectroscopy spectra suggested approximately 2% of $Pr^{4+}$. Assuming that all of the detected $F^-$ and $Pr^{4+}$ content contributed to the electron carrier, the carrier concentration was the lowest in $PrOBiS_2$, higher in $PrO_{1-x}F_xBiS_2$ ($x$ = 0.13), and the highest in $PrO_{1-x}F_xBiS_2$ ($x$ = 0.23). The metallic conduction of all of the compounds can be explained by these electron carriers. The appearance of superconductivity in $PrO_{1-x}F_xBiS_2$ ($x$ = 0.13 and 0.23) can be understood by the high concentration of carrier doping from the high concentrations of F.

Applying pressure is another approach to induce carriers and trigger superconductivity. Figure 5 shows the temperature dependence of the resistance for the $PrOBiS_2$ single crystal under various pressures from 1.7 to 50 GPa. The resistance continued to decrease with the increase of the applied pressure up to 45 GPa, before increasing under 50 GPa. Therefore, up to 50 GPa, the $PrOBiS_2$ single crystal maintained semiconducting behavior under a high pressure and became neither metallic nor superconducting behaviors in the range 2–300 K. Even at 50 GPa, the carrier concentration induced by the high pressure would not sufficient to trigger superconductivity. This semiconducting behavior under applied pressures is inconsistent with the metallic behavior observed in the same crystal without an applied pressure (Fig. 4). Although the reason for this is unknown, the non-hydrostatic pressure via DAC usually degrades the crystallinity of a sample, possibly resulting in mixed transport properties along the *ab*-plane and parallel to the *c*-axis due to the thickness of the boron-doped diamond electrodes.

**4. Conclusion**

F-free PrOBiS$_2$ single crystals were successfully grown by a KCl flux. The XAFS spectroscopy analysis showed that the Pr chemical state had a small amount of tetravalent. The electrical resistivity of the PrOBiS$_2$ single crystals exhibited metallic and semiconducting behaviors above and below 50 K, respectively. The resistance with an applied pressure showed semiconducting behavior in the range 2–300 K. No superconductivity in the PrOBiS$_2$ single crystals was observed down to 0.25 K, or under high pressure up to 50 GPa and down to 2 K.

**Acknowledgments**

The authors would like to thank Dr. K. Fujii (Tokyo Institute of Technology) for useful discussion and valuable advice. The XAFS spectroscopy experiments were conducted at the BL11S2 of Aichi Synchrotron Radiation Center, Aichi Science & Technology Foundation, Aichi, Japan (Experimental No.201801025).

We would like to thank Editage (www.editage.jp) for English language editing.

**Figure captions**

Figure 1. Typical SEM image of a PrOBiS$_2$ single crystal.

Figure 2. XRD pattern of a well-developed plane of a PrOBiS$_2$ single crystal.

Figure 3. Pr L$_3$-edge; XAFS spectroscopy obtained at room temperature for the PrOBiS$_2$ single crystals, Pr$_2$S$_3$ and Pr$_6$O$_{11}$.

Figure 4. Temperature dependence of the resistivity ($\rho$–$T$ characteristics) parallel to the *ab*-plane for the PrOBiS$_2$ single crystal at 1.8–300 K. The inset shows the $\rho$–$T$ characteristics at 0.25–15 K using an ADR option.

Figure 5. Temperature dependence of the resistance in the PrOBiS$_2$ single crystal under various pressures from 1.7 to 50 GPa.

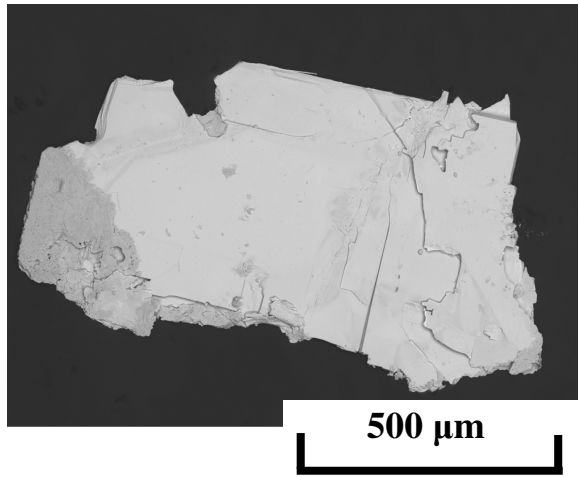

**Figure 1**

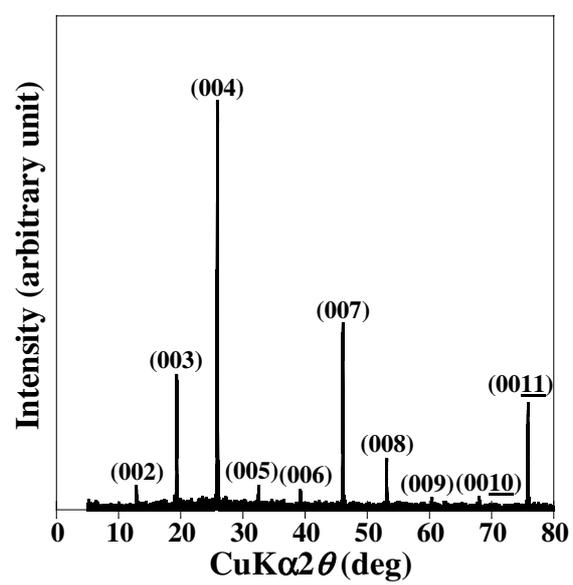

**Figure 2**

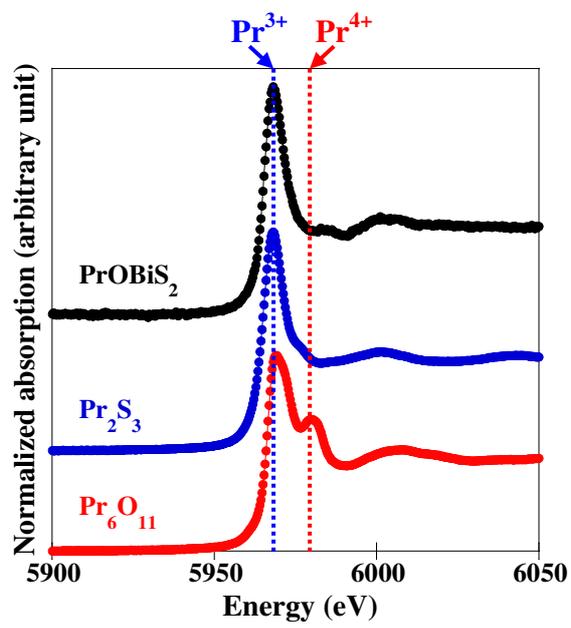

**Figure 3 (Color online)**

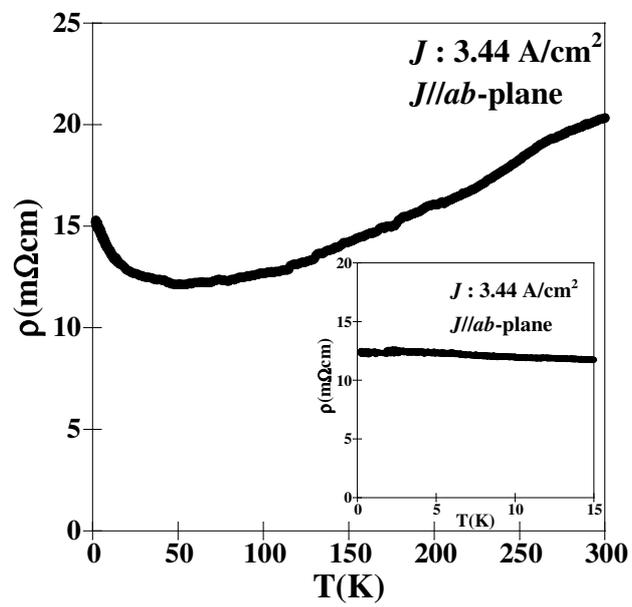

**Figure 4**

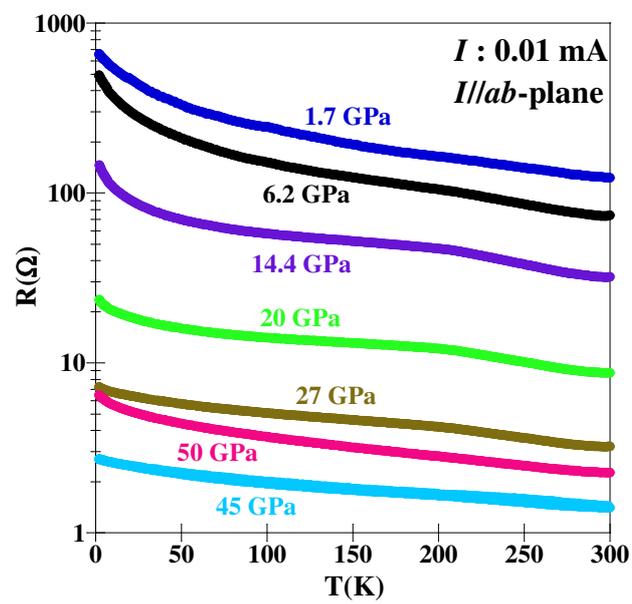

**Figure 5 (Color online)**